\documentclass[fleqn,usenatbib]{mnras}

\usepackage{newtxtext,newtxmath}

\usepackage[T1]{fontenc}

\DeclareRobustCommand{\VAN}[3]{#2}
\let\VANthebibliography\thebibliography
\def\thebibliography{\DeclareRobustCommand{\VAN}[3]{##3}\VANthebibliography}

\usepackage{graphicx}	
\usepackage{amsmath}	

\usepackage{wasysym}    
\usepackage{siunitx}    
\usepackage{booktabs}   

\title[A “New Hope” for Moon Formation]{A “New Hope” for Moon Formation: Presenting a Multiple Impact Pathway}

\author[H. Davies et al.]{
Harrison Davies,$^{1,2}$\thanks{E-mail: h.davies25@imperial.ac.uk}
Philip J. Carter,$^{1}$\thanks{E-mail: p.carter@bristol.ac.uk}
Louis Eddershaw,$^{1}$
Jingyao Dou,$^{1}$
and Zoë M. Leinhardt$^{1}$
\\
$^{1}$H.H. Wills Physics Laboratory, University of Bristol, BS8 1TL, UK\\
$^{2}$Department of Earth Science and Engineering, Imperial College London, SW7 2BP, UK\\
}

\date{Accepted 2025 November 14. Received 2025 November 14; in original form 2025 July 17}

\pubyear{2025}

\begin{document}
\label{firstpage}
\pagerange{\pageref{firstpage}--\pageref{lastpage}}
\maketitle

\begin{abstract}
The leading hypothesis for the origin of the Moon, that of a single giant impact, faces significant challenges. These include either the need for an impactor with a near-identical composition to Earth or an extremely high-mass or high-energy impact to achieve near-complete material mixing. In this paper we explore an alternative, the ``multiple impact hypothesis'', which relaxes the compositional constraints on both the target and projectile, and allows for the consideration of more probable, less extreme impacts that steadily grow the Earth and Moon to their current size over several impact events. Using the hydrodynamical code SWIFT, we simulate ``chains'' of impacts and follow the growth of a moon around a planet analogous to our own. Our results demonstrate that chains of three or more impacts can produce systems comparable to the Earth-Moon system whilst achieving higher compositional similarities than the canonical giant impact scenario. This presents the multiple impact hypothesis as a promising alternative to the single large impact scenario for the origin of the Moon.
\end{abstract}




\section{Introduction}

The majority of natural satellites in our Ssolar system are thought to have either formed alongside their parent planets (such as Io or Titan which orbit Jupiter and Saturn, respectively; e.g. \citealt[][]{CanupWard06}) or were captured when passing too close to their parent body (as is thought to be the case for Neptune's Triton; \citealt{Agnor06}). However, our satellite, the Moon, does not fall into either of these categories \citep{solarsystemmoons}. It is too large to have been captured by the Earth \citep{Tutukov2023} and the lunar material exhibits a depletion in volatiles such as H$_2$O, carbon, and sulfur, which requires formation from highly vaporised material, inconsistent with accretion from a cool protoplanetary disk \citep{Canup2023}. These observational constraints require an alternative theory of formation.

For the last 30 years the prevailing hypothesis for the Moon's origin is that of a single giant impact \citep[e.g.][]{Hartmann1975, Canup2001, Canup2012, Cuk2012,Reufer2012,Lock2018,Carter+20,Kegerreis2022}. One major reason that the giant impact hypothesis is favoured is that a large impact can provide the necessary angular momentum for the current Earth-Moon system. Until recently it was assumed that the angular momentum of the Earth-Moon system was essentially unchanged over the age of the Solar System meaning that if a giant impact was the formation mechanism it would have to both put enough mass in orbit around the Earth to form the Moon and match the angular momentum of the current Earth-Moon system \citep{CanupWC2001}. These requirements result in a narrow parameter space for the impact often referred to as the canonical impact. In this scenario, a projectile of roughly Mars mass collides with the Earth at an oblique angle in the late stages of planet formation and creates a disk of orbiting material \citep[e.g.][]{Canup2004,Nakajima2015,Hosono2019}. In numerical simulations of the canonical impact sufficient material can be ejected to form a disk of approximately 1--2 Lunar masses composed primarily of material originating from the projectile \citep{Canup2004}.

However, the Earth and Moon exhibit near identical isotopic ratios (e.g. calcium, oxygen, silicon, titanium, and tungsten). This suggests that the Earth and Moon formed from identical material sources, with a possible ``late veneer'' accreted onto the Earth explaining the small observed differences \citep{Herwartz2012, Dauphas2014, Young2016}. Hence, in order to match observations, the canonical giant impact scenario requires that the Earth and impactor formed from material with near-identical isotopic ratios. This is very unlikely to happen naturally. Isotopic ratios vary with distance from the Sun, so unless the Earth and the projectile both formed at roughly the same radial distance from the Sun there must be a different explanation for the formation of the Moon \citep[e.g.][]{Kaib2015}. Tungsten also poses particular problems for the canonical model \citep{Kruijer2017}. Tungsten is a siderophile element but W-182 results from the decay of an isotope of the lithophile hafnium, Hf-182; hence, the tungsten isotopic composition of a body ultimately depends on its core formation history. Since the target and impactor are expected to have different core formation histories, it is also expected for there to be a difference between the tungsten isotopic compositions \citep{Touboul07}. 
As a result of these compositional and dynamical constraints the planetary science community has been looking for new solutions to form the Moon. 

A breakthrough came in 2012 when \citet{Cuk2012} found that the angular momentum constraint could be significantly relaxed (interaction with the Sun can sap angular momentum over time, \citealt{Cuk2012,Wisdom2015,cuk2016}) allowing a broader range of impacts to be considered. In particular, impacts that can produce planets and disks with proportionally identical amounts of material from the target and impactor, removing the need for initially (near-)identical compositions. These new impact scenarios, however, rely on a large impactor of similar mass to the Earth  \citep{Canup2012,Reufer2012} or a high-energy, high-angular momentum impact \citep{Cuk2012,Lock17,Lock2018}. Although these new scenarios are consistent with the dynamical and compositional constraints the impacts themselves are uncommon in numerical simulations of planet formation.

\begin{figure}
\includegraphics[width=0.96\linewidth,clip]{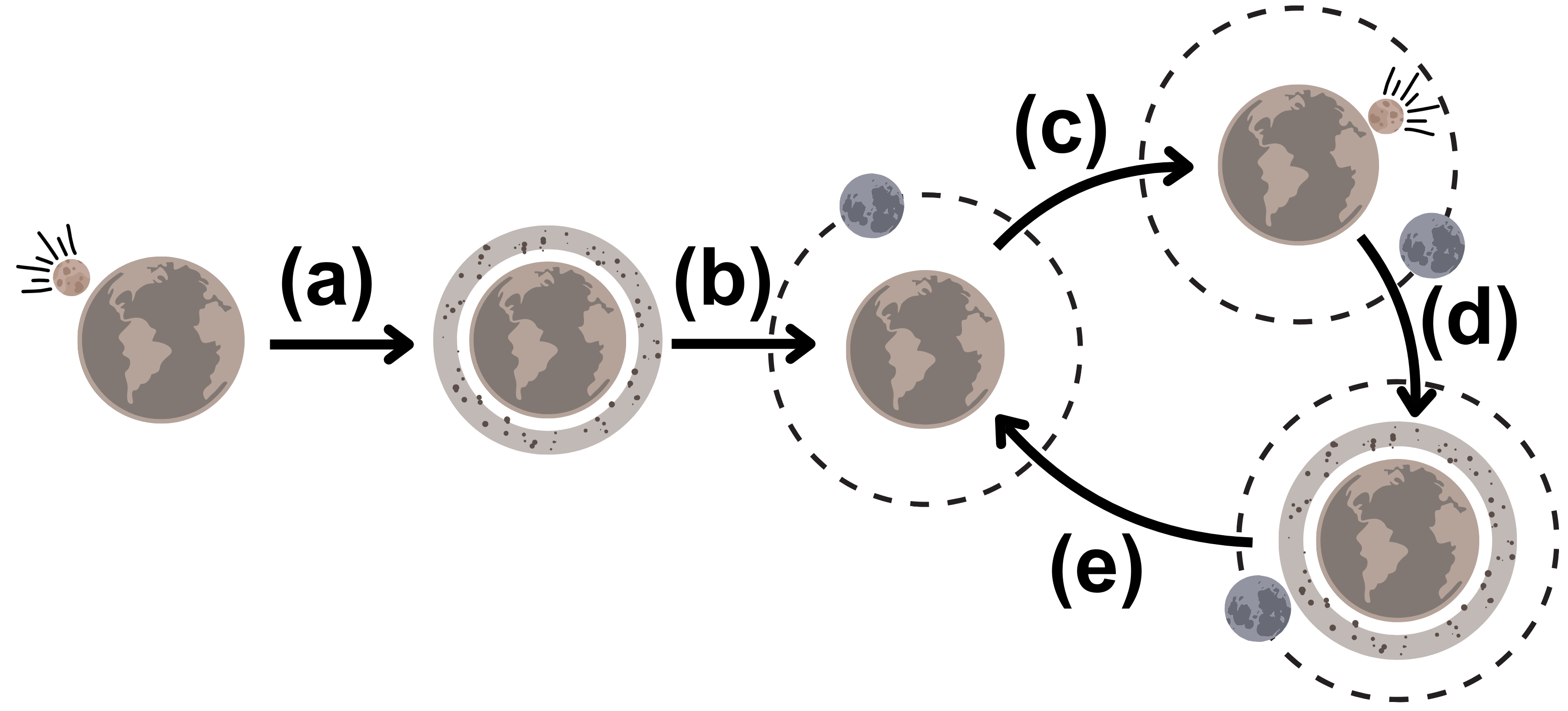}
\caption{Cartoon of the multiple impact Moon-forming scenario: A projectile impacts the proto-Earth forming a debris disk (a), from which an initial seed moon coalesces (b), a subsequent impact occurs (c), forming an additional debris disk (d), from which another moonlet forms and eventually merges with the previously formed moon, and in our case contributes to its mass (e). This process could have repeated until the modern day Moon was formed.} 
\label{multipleimpactchain}
\end{figure}

An alternative model is the ``multiple impact hypothesis'' which replaces the single giant impact with a series of individually less constrained impacts that progressively build up both the Earth and the Moon. Figure \ref{multipleimpactchain} illustrates this process. Previous work by \citet{Rufu2017} investigated a multiple impact scenario involving small, fast impacts that were assumed to have no effect on the target's mass. This work found that, assuming perfect accretionary mergers between moonlets, roughly half of their impact sequences could form moons with a mass comparable to our current Moon after around 20 impacts. This previous work is an excellent step toward consideration of alternative formation pathways for our large satellite. However, we know from numerical simulations of planet formation that planets of terrestrial mass do have several moderate size impacts not just small fast ones. In addition, the impacts themselves will likely change the mass of the growing Earth and the growing satellite may also have some gravitational and dynamical influence on the next collision. Thus, in this paper, we focus on the smaller number of larger impacts, with collision parameters determined by $N$-body simulations of planet formation \citep{Carter2015}, which do appreciably contribute to the mass of the target. Additionally, we consider the effects of intermediate moonlets, as well as the evolving mass and composition of the proto-Earth.

In this paper we simulate ``chains'' of impacts onto rapidly rotating Earth-like targets as they grow from $\sim$ 0.5 to 1.0\,M$_\oplus$ over four separate impact events. We compare the end result of each chain against the mass, angular momentum, and compositional constraints of the current Earth-Moon system to assess the viability of this hypothesis compared to a single giant impact (canonical, high-energy, and equal mass). 

\section{Methods}\label{methods}

We follow the development of a planet-satellite system over a ``chain'' of four impacts. Each impact is a distinct simulation with initial conditions determined by the final state of the previous simulation. The accrued masses of the planet and disk dictate the bodies that are selected for simulation in the next chain ``link''. A set of discrete target masses ranging from 0.57--1.28\,M$_\oplus$ were used as initial target masses in each simulation.

The initial mass of Earth when it began experiencing moderate-mass impacts is not well constrained. However, solar system planet formation simulations \citep{Carter2015} show that planets that eventually grow to Earth mass experienced their first moderate-mass (\textgreater 0.04\,M$_\oplus$) impacts when they reached between 0.57 and 0.66\,M$_\oplus$. Based on these findings, we set the target planet's initial mass for the first chain-link at 0.569\,M$_\oplus$ (see \S\ref{sec:initial}). All targets were given the same initial rotation period of three hours matching the fastest rotation used in \citet{Rufu2017}. For an Earth mass planet, this period corresponds to half of the tidal-breakup spin. Faster spins (smaller rotation periods) allow mass to be ejected from the surface of the planet more easily, increasing the mass of the disk. However, for high impact angular momentum the disk mass is limited by ejection from the system \citep[see][]{Rufu2017}.

After each impact the disk that orbits the planet is analysed and its mass is converted into an equivalent ``moonlet'' mass that would be expected from a disk of that size (see \S \ref{sec:postimp}). The moonlets are simulated as single particles carrying the entire moonlet mass, and these interact purely gravitationally (not hydrodynamically) with the other particles in the simulation. The moons were also placed with an orbit at a fixed distance from the planet, at 8\,R$_\oplus$ and the spin of the planet is reset to three hours. In our simulations with a mean final target mass of 0.87 M$_\oplus$, this results in a mean final target angular momentum of 1.11 times the current angular momentum of the Earth-Moon system. This setup means that angular momentum is not conserved between impacts in the chain. The angular momentum introduced by each impact in the chain is not carried forward in full, which likely leads to a lower angular momentum growth over the chain than expected. We chose to adopt this simplification due to the lack of constraints on the tidal parameters and their evolution during the formation of the Moon.

The possible masses of impactors were randomly sampled from a distribution of likely object masses for impacts onto Earth-like planets. This distribution was based on previous $N$-body simulations of planet formation \citep{Carter2015}, and was constrained to select only impactors with masses of at least 0.04\,M$_\oplus$, below which our simulations did not produce debris disks capable of forming appreciable moonlets. To model the impacts we used the smoothed particle hydrodynamics (SPH) code SWIFT \citep{SWIFT}, configured for planetary hydrodynamics and equations of state, selecting a cubic spline kernel, with a maximum smoothing length, $h_\mathrm{max}$, of 0.2\,R$_\oplus$, and a total particle count of around 600,000. This choice of $h_\mathrm{max}$ sets a density floor at approximately 2.2\,\si{\kilogram \per \meter \cubed}. 

\subsection{Initial Conditions}\label{sec:initial}

The impactors and targets in each chain-link were constructed as two-layer SPH planets using WoMa \citep{WoMa}. The planets were made with a 30\% by weight ANEOS iron core \citep{StewartANEOSIron2020} with a 70\% by weight ANEOS forsterite mantle \citep{Stewart2020}. Each layer's temperature-radius profile was determined using a power law relationship with density, $T \propto \rho^2$, with a fixed surface temperature of 2000 \si{\kelvin} and fixed surface density of 3300 \si{\kilogram \per \meter \cubed}. We used a fixed particle mass between each body, scaling the number of particles with the mass of the body, resulting in a total particle count of approximately 600,000 per simulation. We constructed 33 different impactors ranging from 0.025 to 0.185\,M$_\oplus$, alongside 70 planets ranging between 0.569 and 1.28\,M$_\oplus$. Subsequent chain links selected the nearest available target to the determined planet mass resulting from the previous impact, from the pool of 70. By assuming rigid body rotation for the stable pre-impact target, the targets were given a rotational period of three hours, corresponding to about 49\% of the rotational breakup angular velocity. Impactor masses were randomly drawn from a distribution approximated from $N$-body simulations of solar system formation \citep[simulation numbers 21 and 23 from][]{Carter2015}. The distribution consisted of four mass bins weighted by the frequency of those impacts within the simulation, with values selected uniformly within each bin. WoMa was then used to establish impact conditions with an initial impactor trajectory such that the impact would occur at an angle of 45\si{\degree} (an impact parameter of 0.707), corresponding to the most probable impact angle, with a speed equal to the mutual escape speed of the target-impactor system. We chose to limit our simulations to the most likely impact angle and a likely impact speed in order to reduce the possible parameter space for computational efficiency. The initial conditions were set such that the impacts occurred in the prograde direction, aligned with the target’s initial rotation. The particles are tagged such that the origin of particles, later found in the disk or resultant planet (impact ``remnant''),  could be monitored.

\subsection{Post Impact Analysis}\label{sec:postimp}

Each chain-link was simulated for 37.5 hours, including 1.5 hours leading up to the moment of impact. Once completed, the simulation was analysed and the next moonlet mass in the chain was predicted from the resulting debris disk. The centre of the remnant planet was deemed to be the particle with the lowest gravitational potential, which is then used as the origin of a radial density profile. By smoothing the density profile slightly, the location where the density profile dips below a threshold value of 250 \si{\kilogram \per \meter \cubed} was then quoted as the radius of the remnant, and the remnant mass as the sum of the particle masses within that radius. This density threshold was identified as an effective cut-off for distinguishing the rotating planetary surface from the lower density orbiting disk. Using this remnant's mass and position, the approximate orbital trajectories of every particle in the simulation were calculated. Particles were categorised as ``unbound'' if they had positive specific orbital energy, and categorised as disk material if the orbital periapsis was greater than the remnant planet's radius (i.e. if they were on a non-colliding orbit). The total mass and angular momentum of the disk were then calculated and used to predict the resulting moonlet mass using the semi-analytical relationship developed by \citet{Salmon2012}, who simulated the disk evolution and moonlet formation in comparable fluid disks:
\begin{equation}
    M_{moonlet} = 1.14\frac{L_d}{\sqrt{GM_\oplus a_R}}-0.67M_d-2.3M_\infty,
\end{equation}
where $L_d$ and $M_d$ are the angular momentum and mass of the disk respectively, $a_R$ is the Roche radius of the planet and $M_\infty$ is the mass of material that escapes during the moonlet's accretion. This equation assumes moonlet formation at a distance of 2.15 $a_R \approx $ 3.2 R$_\oplus$, which corresponds to the average formation radius observed in their simulations. Additionally, we adopt an escaped mass fraction of $M_\infty/M_d = 0.05$, chosen as a conservative estimate that falls within the upper range of values found within their simulations, in order to avoid overestimating the moonlet mass. We note that this relation does not take into account any effect a pre-existing moonlet may have on disk evolution.

The planetary mass that is fed into the next chain-link is calculated as
\begin{equation}
    M_{next\,planet} = M_{bound} - M_{moonlet} - M_\infty,
\end{equation}
and this analysis is repeated for each of the four chain links.

\subsection{Compositional ``Distance''}
By identifying the origin of each particle in the planet and disk at the end of a simulation, we can track the changing composition of the target and moon over the entire chain. To quantify the compositional difference between a resulting Earth and Moon, a deviation percentage is often used,
\begin{equation}\label{distance}
     \delta f_T \equiv  \left ( \frac{F_{D, tar}}{F_{P, tar}} - 1\right ) \times 100
\end{equation}
where $F_{D, tar}$ and $F_{P, tar}$ are the fractions of silicate material originating from the target that end up in the disk and planet respectively \citep{Canup2012}. However, since the planets and moons formed in this paper are composed from five different material sources (one initial planet + four impactors), this formula does not fully capture the deviation between their material populations. While the planet and disk may share the same proportions of their mass from the target, their remaining mass could come from very different proportions of each impactor, increasing their observed compositional differences. Therefore, we define a compositional ``distance'' between the material distributions of the two bodies
\begin{equation}
    d_c \equiv \sqrt{\sum^{N}_{i}(F_{D, i} - F_{P, i})^2}
\end{equation}
where $F_{D, i}$ and $F_{P, i}$ are the fractions of silicate material originating from the specific impactor $i$, for the disk and planet, respectively, and $N$ is the number of impactors or chain-links. Here, a distance of 0 describes two bodies with proportionally identical contributions from each material source, and a larger distance indicates a greater disparity in their compositions. As the number of impactors is increased, the potential maximum distance increases as $d_{c, max}  = \sqrt{N}$.
Table \ref{DEVIATIONDISTANCECOMPARISON} shows the deviation percentages of a range of simulations from this and previous works \citep{Canup2012,Reufer2012}, and the corresponding compositional distances.

\begin{table*}
\caption{Compositional properties of various impact simulations. Shown are the initial impactor masses ($M_i$), the deviation percentages ($\delta f_T$), the silicate fractions of the disks from the targets $F_{D,T}$, and the compositional distances ($d_c$) after one impact and after four impacts.}
\begin{tabular}{@{}lccccc@{}}
\toprule
Source & $M_{i}$ ($M_\oplus$) & $\delta f_T$ (\%) & $f_{D,T}$ (\%) & $d_c$ 1st impact  & $d_c$ all impacts \\
\midrule
Chain 1\footnotemark[1] & 0.12 & -55 & 38 & 0.47 & 0.238 \\
Chain 4\footnotemark[1] & 0.08 & -71 & 26 & 0.64 & 0.273\\
Chain 8\footnotemark[1] & 0.09 & -74 & 23 & 0.66 & 0.312\\

cA08\footnotemark[2] & 0.1 & -66 & 31 & 0.60 & \\
cB04\footnotemark[2] & 0.15 & -41 & 53 & 0.37 & \\
cC03\footnotemark[2] & 0.2 & -37 & 54 & 0.32 & \\
cC04\footnotemark[2] & 0.2 & -32 & 58 & 0.27 & \\
cC05\footnotemark[2] & 0.2 & -36 & 54 & 0.30 & \\
cC06\footnotemark[2] & 0.2 & -35 & 56 & 0.30 & \\
cC07\footnotemark[2] & 0.2 & -54 & 39 & 0.46 & \\
cC08\footnotemark[2] & 0.2 & -76 & 20 & 0.63 & \\

Run 32\footnotemark[3] & 0.47 & -8 & 51 & 0.05 & \\
Run 33\footnotemark[3] & 0.47 & -11 & 49 & 0.06 & \\
Run 35\footnotemark[3] & 0.47 & -5 & 52 & 0.03 & \\
Run 43\footnotemark[3] & 0.47 & -15 & 48 & 0.08 & \\
Run 7\footnotemark[3] & 0.42 & -7 & 56 & 0.04 & \\

\bottomrule
\end{tabular}
\label{DEVIATIONDISTANCECOMPARISON}%

\footnotemark[1]{This paper.}\\
\footnotemark[2]{$\sim$10-20\% Earth mass impacts, \citet{Reufer2012}.}\\
\footnotemark[3]{\textgreater 40\% Earth mass impacts, \citet{Canup2012}.}
\end{table*}

\section{Results}

\begin{figure*}
\includegraphics[width=0.98\linewidth,clip]{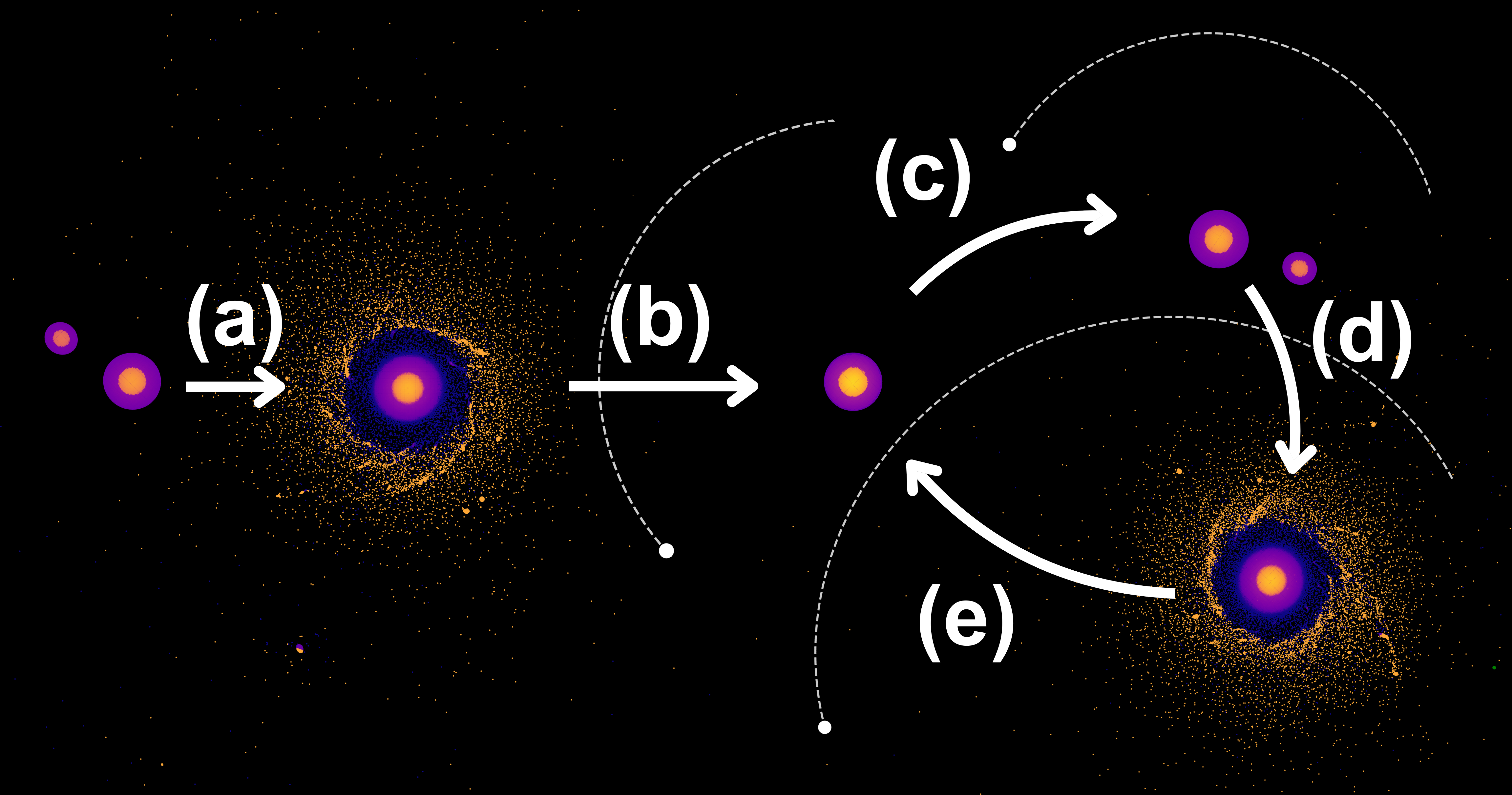}
\caption{Progression of the first two links of Chain 1. Illustrating an initial disk forming impact (a) from which we create a larger planet and an initial moonlet (b), to a subsequent impact (c) which forms another debris disk (d), from which we again create a larger planet and add to the mass of the moonlet, before repeating the process for the next chain link. The white particle with the trailing dashed line shows the point mass moon and a segment of its instantaneous orbit, the green particles indicate unbound material, and the orange particles represent the disk material. The remaining particles are coloured by their density with yellow indicating the most dense material. The images present a slice taken along the centre of mass.} 
\label{imagechain}
\end{figure*}

Using the SPH software SWIFT \citep{SWIFT}, we simulated 12 impact ``chains'' each consisting of four individual impacts drawing impactor masses from a distribution based on previous $N$-body simulations \citet{Carter2015}. Conventional impact parameters were selected, with the impact speed set to the mutual escape velocity, $v_{esc}$, and an impact angle of 45\si{\degree}. The main properties characterising the outputs of these simulations are presented in Table \ref{resultstable}. These include the final planet mass ($M_p$) and moon mass ($M_m$) both expressed as fractions of their modern values, the iron-to-total mass ratio of the disk ($R_{Fe}$), and the total angular momentum of the system, including the planet, moon, and disk ($L_{p,m,d}$) as a fraction of $L_{EM}$, the current angular momentum of the Earth-Moon system. We also measure a compositional ``distance'' to quantify the expected similarities between the two bodies. This is calculated as the Euclidean distance between the source compositions of the two bodies (see Section \ref{methods}). Snapshots taken from the first two impacts of Chain 1 are shown in Figure \ref{imagechain}, demonstrating the progression of our collision chains.

\begin{figure}
\centering
\includegraphics[width=0.96\linewidth,clip]{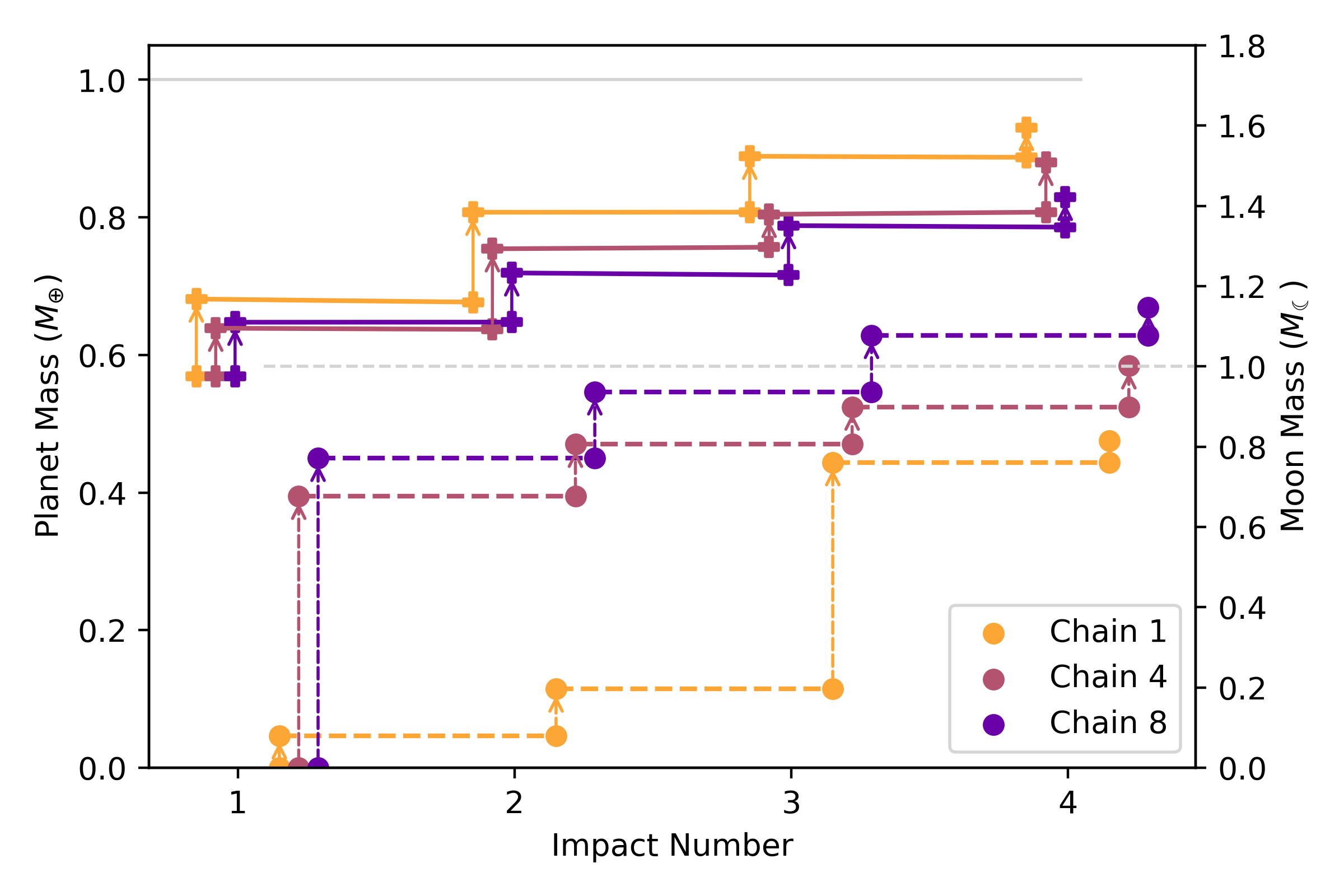}
\caption{The accumulated mass of the moonlet (dashed lines) and of the planet (solid lines) over the course of the four impacts in Chains 1, 4, and 8. The grey lines at a value of 1.0 on their respective axes indicate the masses of the modern Earth and Moon.} 
\label{massplot}
\end{figure}

\begin{figure}
\centering
\includegraphics[width=0.96\linewidth,clip]{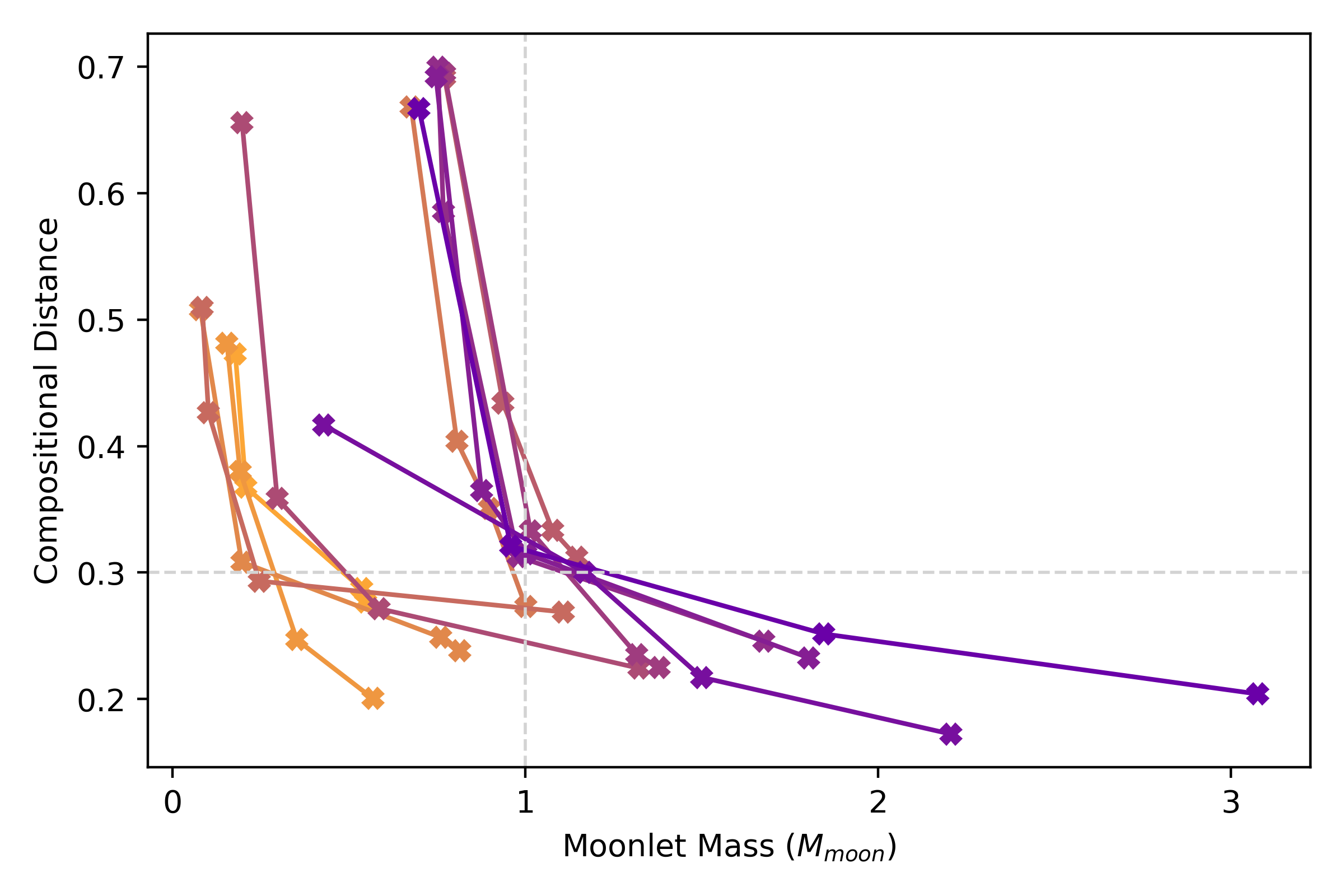}
\caption{The compositional distance between the planet and moon, as defined in Equation \ref{distance}, over the accrued moon mass for each of the 4 links within the 12 chains. The two grey lines indicate the ``successful'' thresholds for the Distance and Mass at 0.3 and 1 respectively, with those chains ending in the lower-right-hand quadrant considered successful in this regard. Individual chains are coloured from orange to purple, ordered by increasing final moonlet mass.}
\label{distancevsmass}
\end{figure}

All chains produced moonlets with final masses $\geq$ 0.55 M$_{\leftmoon}$ (the modern mass of the Moon) and with core-to-total mass ratios, $R_{Fe}$, from 2.56--6.02\%. These systems formed with a range of compositional distances, $d_c$, between the planet and moon, from as low as 0.172 up to 0.312. The final system angular momentum also varied significantly from 1.71--3.1 $L_{EM}$. The results from all chains are presented in Table \ref{resultstable}. The evolution of the planet and moonlet masses for chains 1, 4, and 8 are presented in Figure \ref{massplot}. Additionally, Figure \ref{distancevsmass} displays the relationship between the compositional distance and moonlet mass over the chain for all simulated collision chains. The initial impacts typically produce a sub-lunar moonlet with a substantial compositional ``distance'' from the target. The compositional distances calculated after the first impact in each chain are similar to the compositional distance calculated for a previous simulation of the canonical Moon-forming impact (see Table \ref{DEVIATIONDISTANCECOMPARISON}). However, as the chains progress, the target and moonlet, both growing from the same impactors, become increasingly similar in their compositions.

\begin{table}
\caption{Properties of the impact chains. The final planet ($M_p$) and moon ($M_m$) masses in units of their modern values are shown alongside the disk iron-to-total mass ratio ($R_{Fe}$), the compositional distance ($d_c$), and the total angular momentum of the planet moon and disk ($L_{p,m,d}$) in units of the angular momentum of the modern Earth-Moon System ($L_{EM}$).}
\centering
\begin{tabular}{@{}lrrrrrr@{}}
\toprule
Chain & $M_p$ & $M_m$ & $R_{Fe}$ & $d_c$ & $L_{p,m,d}$\\
 & ($M_\oplus$) & ($M_{\leftmoon}$) & (\%) &  & ($L_{EM}$)\\
\midrule
1 & 0.9306 & 0.81 & 4.21 & 0.238 & 1.7514 \\
2 & 1.0676 & 2.21 & 2.75 & 0.172 & 2.7618 \\
3 & 1.0158 & 1.38 & 2.61 & 0.225 & 2.2339 \\
4 & 0.8799 & 1.00 & 4.26 & 0.273 & 1.8500 \\
5 & 0.9980 & 0.57 & 6.02 & 0.200 & 2.1483 \\
6 & 0.9369 & 1.32 & 2.57 & 0.224 & 2.1702 \\
7 & 0.9750 & 1.11 & 4.56 & 0.269 & 2.2959 \\
8 & 0.8290 & 1.15 & 2.97 & 0.312 & 1.5810 \\
9 & 0.9118 & 0.55 & 2.56 & 0.277 & 1.7074 \\
10 & 0.9857 & 1.80 & 2.80 & 0.233 & 2.6845 \\
11 & 1.0244 & 3.08 & 2.67 & 0.204 & 3.0674 \\
12 & 0.9871 & 1.68 & 2.64 & 0.246 & 2.6205 \\
\bottomrule
\end{tabular}
\label{resultstable}
\end{table}

Nine of the chains were able to form moons as large as the Moon and also resulted in planets with masses in the range of 0.829 to 1.068 $M_\oplus$. The compositional distance ($d_c$), measured between each of these pairs of planets and moons, varied from 0.172 to 0.312 and $R_{Fe}$ varied from 2.57 to 4.56\%, with total final system angular momenta ($L_{p,m,d}$) that ranged between 1.59 and 3.10 $L_{EM}$. The angular momentum presented here should likely be considered a lower limit, given the simplifications discussed in section \ref{methods}. One of these chains (chain 4) formed a moon of 1 lunar mass in a system with an angular momentum within the allowable range of 1--2 $L_{EM}$ (having sufficient angular momentum to match the current value, or allowing the evection resonance to potentially explain its evolution to the current observed value, \citealt{Cuk2012}). Additionally, this same chain produced a moon with an easily sufficient depletion in iron (less than 5\% by weight \citealt{Canup2004}) to account for the Moon's small iron core. This chain did however produce a low-mass Earth at only 0.88\,$M_\oplus$.

\section{Discussion}

We consider a successful Moon forming chain to be one that meets the following constraints:
\begin{enumerate}
    \item A central remnant no larger than 1\,M$_\oplus$.
    \item A satellite at least as large as the modern Moon at 1\,M$_{\leftmoon}$.
    
    \item A satellite with an iron content of \textless 5\% \citep{Canup2004}.
    \item A compositional distance $\lesssim$0.3.
    \item A total system angular momentum $\lesssim$\ 2\,$L_{EM}$ \citep{Cuk2012}.
\end{enumerate}
We choose the planet-moon compositional distance requirement to result in a matching or improving upon the similarities produced by the canonical and slow hit-and-run \citep{Reufer2012} single giant impact models. Our conservative angular momentum requirement is based on the maximum value that has been demonstrated to be removable via the evection resonance or Laplace plane transition \citep{Cuk2012, Cuk2021}

Looking at this wider selection of ``success” criteria compared with previous work \citep{Rufu2017}, we find that two out of our 12 simulated impact chains produced systems that are consistent with the modern Earth-Moon system. Only nine of the chains produced a satellite as large as the Moon, and three of those produced planets more massive than the Earth. The iron mass content requirement is met by all but one chain.
Specifically, chains 4 and 8 fulfil all of these criteria while producing compositional distances competitive with the best possible within the canonical single giant impact models (0.273 and 0.312, see Table \ref{DEVIATIONDISTANCECOMPARISON}).

Chain 4 is considered our best outcome, successfully recreating the lunar mass, angular momentum, and moon core ratio, alongside a low compositional distance, while only falling short with a lower-than-desired planet mass. However, we have only considered accretion via (moderate) giant impacts and it is expected that the Earth would have gained a significant mass from smaller bodies as well during this time period \citep[e.g.][]{Carter2015}. Additionally, we assume that all moonlets in our simulations merge, which may not always be the case \citep{Citron18}. Gravitational interactions between moonlets could cause smaller moonlets to scatter, either into the planet or out of the system, before any moonlet collision occurs; or such collisions could eject some of the moonlet mass. As such, it is also plausible that additional moderate-mass impacts could have occurred, contributing to the target's mass without significantly affecting the moon's formation.

While these simulations fulfil our criteria, it is important to note that the angular momentum recorded at the end of our simulation chains is not a robust measure of the true angular momentum at the conclusion of lunar formation in these scenarios. In our simulations, the angular momentum is effectively reset between each impact (the new value being set by the mass of the new moonlet and planet at the fixed orbital radius and spin period, in the initial conditions for the final impacts our Earth-Moon angular momentum averaged 1.35 $L_\mathrm{EM}$), likely reducing the accrued total angular momentum after all but the final impact. As a result, while the angular momentum increases between impacts due to the growing mass of the moon, alongside the mass and radius of the planet, the ``final'' angular momentum presented here may underestimate the true angular momentum that would be produced in a system with a similarly high initial angular momentum. On the other hand, considering a wider range of collisions, e.g.\ with various impact angles (rather than at a fixed angle of 45\si\degree) and including retrograde collisions, would likely lead to a lower overall angular momentum. Hence, the true angular momentum could plausibly be higher or lower than the values listed in Table \ref{resultstable}. Angular momenta up to $\sim$2.4\,$L_\mathrm{EM}$ have been shown to be lost through resonant interactions over time, see e.g.\ \citet{Cuk2012}, \citet{Wisdom2015}, \citet{cuk2016}, and \citet{Cuk2021} though the applicability of these mechanisms to the multiple-impact scenario is unclear. No upper limit for angular momentum removal has been demonstrated and so it is possible that the present angular momentum of the Earth-Moon system could be achieved for all of our simulation chains.

It is important to note that the mechanisms for removing excess angular momentum from the Earth-Moon system have not been explicitly tested for the multiple impact scenario. For the evection resonance, the Moon is typically captured into resonance quickly after formation compared to the expected timescale between impacts \citep{Cuk2012}. In order to remove substantial angular momentum in the multiple impact model, it would thus likely be necessary for a new moonlet to be captured into resonance in a system that already has an existing moonlet. It is not clear in this scenario whether capture is possible or whether the efficiency of angular momentum removal would be comparable to the single impact scenario. For the Laplace plane transition to remove substantial angular momentum the Earth is required to have high obliquity (>60\si\degree) following the formation of the Moon \citep{Cuk2021}. For a random distribution of impact directions, it is unlikely that most of the multiple impacts required to grow the Moon in this scenario would push the Earth's obliquity in the same direction. Since impactors are generally low mass compared to the proto-Earth, it is also unlikely that a single event in the chain could produce such a high obliquity. A high obliquity may therefore be required before the chain of impacts that leads to the formation of the Moon in order for the transition to extract sufficient angular momentum. Future investigations of the multiple-impact pathway to Lunar formation should consider the angular momentum and obliquity evolution in more detail.

In our simulations, we have assumed that the time between impacts is much longer than the time required for accretion and collision of moonlets and their tidal evolution to a distance of 8 R$_\oplus$. The typical intervals for giant impacts, or the smaller impacts we study here, during the later stages of terrestrial planet growth are Myrs to 10s of Myrs \citep[e.g.][]{Quintana16,Carter22}. The expected time between impacts is thus much longer than the months-to-years timescale for moonlet accretion \citep{Kokubo2000,Lock2018}, and longer than the 10s to 100s kyr timescale \citep{Cuk2012} for tidal evolution to raise the Moon's orbit to $\sim$8\,R$_\oplus$ (although it should be noted that the tidal evolution for moonlets would differ from that of a fully grown Moon). \citet{Citron18} found that the typical timescale for moonlet collisions is <0.5\,Myr, though this may be sensitive to the assumed interval between impacts and hence tidal recession timescale. The typical interval between impacts is longer than the other important evolutionary timescales. Moonlets may have reached larger semi-major axes than used in our simulations in the time between subsequent impacts, but this is unlikely to substantially change our results.

In addition to the impact chain simulations, we ran several preliminary test simulations of individual impact events in the presence of an existing moon. These simulations used the same setup for target and impactor as described in section \ref{methods} and included a hydrodynamic lunar mass moon placed in a circular orbit at 10\,R$_\oplus$ from a non-rotating Earth-mass target. These preliminary simulations indicated that a large pre-existing moonlet significantly influenced the disk formed by a subsequent impact onto the proto-Earth, primarily increasing the mass of the disk, in contrast to the findings of \citet{Rufu2017} from their test simulations\footnote{Note that \citet{Rufu2017} did observe a significantly different outcome if their much more massive pre-existing moon orbited at $\sim$8.7\,R$_\oplus$ or less.}. The presence of a moonlet had the smallest effect on the disks produced by the highest mass impactors and so would likely have had little effect on the 0.1\,M$_\oplus$ impactor tested by \citet{Rufu2017}. However, disks produced by smaller impactors from 0.005 to 0.015\,M$_\oplus$ were affected significantly by the presence of the moonlet, with disk masses increasing by 4\% to 35\% compared to the no moon case and averaging 19\%. The largest increases in mass were observed when the moonlet passed through the plume of ejected material. All our simulations additionally demonstrate that the impacts have a significant effect on the moonlet’s orbit, consistently raising or lowering it, depending on the moonlet’s position in its orbit at the time of the impact. When the moon’s initial velocity was aligned with the impact the semi-major axis was reduced and when anti-aligned it was increased, with the strength of this effect correlating with the impactor mass, and ranging from -2.5 to +3.3\,R$_\oplus$. These effects on both the disk and moonlet are not insignificant, potentially influencing the moonlet’s stability while also increasing the rate at which moon-forming material could be accumulated. Consequently, the presence of pre-existing moonlets should be considered in studies of subsequent impacts.

The successful formation of moonlets with both total and core mass comparable to the current Moon, alongside planets with masses comparable to the Earth, demonstrates that a series of sequential impacts could have produced the modern day Earth-Moon system. The resultant Earth and Moon masses fluctuate between the simulations presented here, but they appear to indicate a promising convergence near the required values.

Table \ref{DEVIATIONDISTANCECOMPARISON} compares the results of this paper with compositional distances from previous work, showing that the distances for chains 4 and 8, around 0.30, improve upon those achievable within a canonical giant impact scenario (simulation cA08 from \citealt{Reufer2012}). Additionally, our compositional distances are in line with those produced by larger impact models using impactor masses greater than 0.2\,$M_\oplus$. The most extreme impact models (`half-Earth' impacts \citealt{Canup2012}) can produce compositional distances as low as 0.03, and other synestia-forming collisions \citep{Lock2018} may be able to improve upon this further. However, our results are achieved with much smaller impacts, which are more common during the formation of Earth-like planets than those considered in these more extreme scenarios.

Mixing material from multiple different bodies may allow isotopic differences to be reduced compared to the canonical Moon-formation scenario. One of the biggest problems for the canonical model is tungsten \citep[e.g.][]{Kruijer2017} because the isotopic composition is controlled by core-formation history rather than formation location. The near-identical tungsten isotopic composition of the Earth and Moon seemingly requires a high degree of mixing during the formation of the Moon. The multiple impact scenario allows for greater mixing (lower compositional distance) than the canonical scenario, but may also allow greater preservation of heterogeneity within the Earth and perhaps the Moon as well \citep{Rufu2017}. The additional core-formation events on the Earth inherent with the multiple impact model also provide further opportunities to alter the tungsten composition of Earth's mantle. With a single impact origin for the Moon, if there is not near-complete mixing or near-identical initial tungsten isotopes, the Earth and Moon would have different tungsten isotopic anomalies. However, with multiple impactors that could potentially possess initial tungsten anomalies offset in opposite directions relative to the proto-Earth, it is possible for the final averaged anomaly for the Moon to be closer to that of the Earth than our compositional distance measure would suggest. Whether an appropriate combination of mixing and core formation histories for each impactor to produce such a result is likely, however, is difficult to assess.

By considering a rapidly rotating target, we have shown that the same impacts expected to have accumulated Earth's mass, consistent with $N$-body simulations of planet formation \citep{Carter2015}, could also have incrementally built up the mass of the Moon. This approach also reduces the level of compositional similarity required between the target and impactors over the canonical single giant impact, overall indicating a more probable scenario for the formation of the Earth and Moon.

\section*{Acknowledgements}

This work was carried out using the computational facilities of the Advanced Computing Research Centre, University of Bristol - http://www.bristol.ac.uk/acrc/. 
PJC and ZML acknowledge financial support from the UK Science and
Technology Facilities Council (grant numbers: ST/V000454/1 and
ST/Y002024/1).

\section*{Data Availability}

The data supporting these findings, including simulation outputs and analysis scripts, are available from the corresponding authors upon reasonable request.



\bibliographystyle{mnras}
\bibliography{bibliography} 




\bsp	
\label{lastpage}
\end{document}